\documentclass{article}
\usepackage[utf8]{inputenc}
\usepackage{graphicx}
\usepackage{authblk}
\usepackage{lineno}
\usepackage[margin=1in]{geometry}
\usepackage{url}
\usepackage{listings}
\usepackage{amsmath}

\title{Moon and background removal algorithm for all-sky imager}
\author[1]{S. Tian}
\affil[1]{Department of Atmospheric and Oceanic Sciences, University of California, Los Angeles, California, United States}
\date{Contacting email: ts0110@atmos.ucla.edu}

\begin{document}

\maketitle

\begin{abstract}
All-sky imagers (ASIs) are used to record auroral activities from the ground but are often contaminated by the moon. Here, we studied the THEMIS ASIs data and developed an algorithm to eliminate the moon which can be generalized to other types of ASIs. With our algorithm, the ASI pixels within the moon's surface are typically saturated and thus removed by the algorithm. The ASI pixels within the moon's glow are close to but not saturated and thus can be calibrated by the algorithm to recover auroral structures within the glow. For pixels far away from the moon or when there is no moon, the algorithm preserves typical auroral forms, from the transient features of auroral streamers and pulsating aurora to more stable features of pre-onset arcs. Note that the algorithm does not treat clouds, which is a known limitation.
\end{abstract} 

\section{Introduction}

\begin{figure}[h]
\centering
\includegraphics[width=6in]{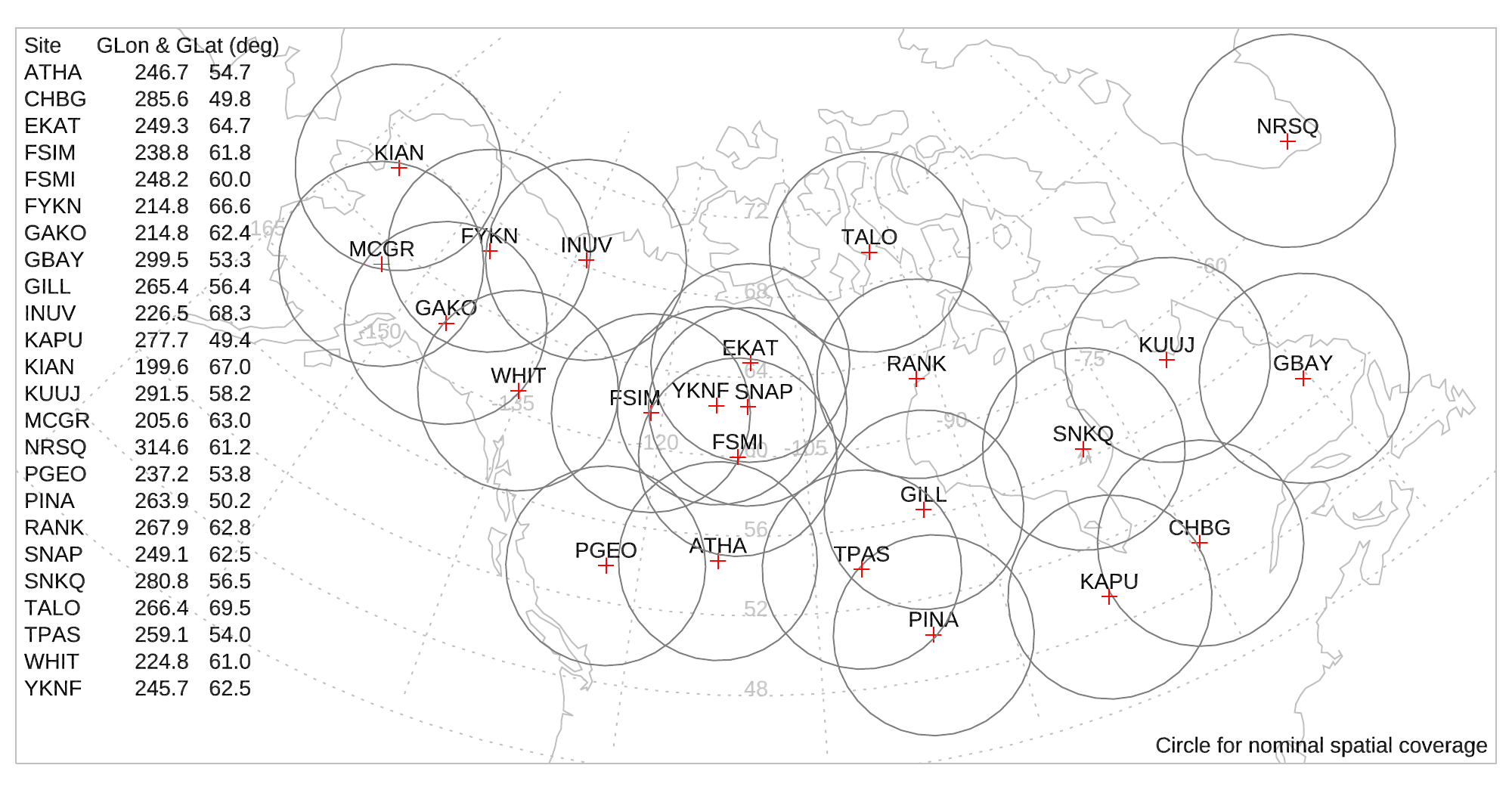}
\caption{The site map marks the location and nominal spatial coverage of the THEMIS ASIs over the northern American continent. Adopted from \cite{mende_2008_asi}.}
\label{fig:site_map}
\end{figure}

All-sky imagers (ASIs) are used to record auroral activities from the ground but are often contaminated by the moon. Here, we study the THEMIS ASIs \cite{mende_2008_asi} to eliminate the effect of moon and its glow. The THEMIS ASIs are fish-eye cameras record auroral images at a cadence of 3 sec. Each frame is a image of $256\times 256$ pixels, covering a field of view of $\sim 2\pi$ solid range above $\sim 0$ deg elevation angle. Each pixel has a raw CCD count ranges from 0 to 65535, characterizing the brightness or the number of photons. The pixels on the moon's surface typically reach the saturation count (65535). The THEMIS ASIs typically operate many hours continuously per day during the night times. The full-resolution images are saved in the CDF format in every 1 hour. The data are publicly accessible at \url{http://themis.ssl.berkeley.edu/data/themis/thg/l1/asi/}. The THEMIS ASIs contain a network of imagers over the northern American continent. Fig \ref{fig:site_map} shows the site map marks the location and nominal spatial coverage of these imagers. In the rest of this report, we will focus on the full-resolution images ($256\times256$ pixels) from these sites. We will introduce the algorithm and some key technical details in the following sections.

\section{Algorithm Overview}
\label{sec:overview}

As mentioned in the previous section, for a certain ASI site, its raw images $r_{m,m}[n]$ is an array in $n,m,m$. It has $n$ frames and each frame is an $m\times m$ image, where $m=256$ and $n$ is typically larger than $6,000$, which corresponds to more than 5 hours at a cadence of 3 sec. In the rest of the paper, we will use the iterators $k$ for looping in frame (or time) and $i$ and $j$ for looping through the pixels at each frame. For example, we use the notation $r_{m,m}[n]$ to refer to the array of $[n,m,m]$, when the actual dimensions $n$ and $m$ are used. To refer to a subset of $r_{m,m}[n]$, we use $r_{i,j}[n]$ to refer to the array of $[n]$, where the iterators $i$ and $j$ are used to specify a certain pixel in $[0,m)$. Similarly, we use $r_{m,m}[k]$ to refer to the array of $[m,m]$, where the iterator $k$ specifies a certain frame in $[0,n)$. 

Traditionally, one would remove the moon by analysing the image at each frame ($r_{m,m}[k]$) and loops through all frames ($k$). Here, our algorithm analyzes $r_{i,j}[n]$, which is the time series of the count at a certain pixel $(i,j)$ and loop through all pixels ($i$ and $j$). To see why it works, let us see how $r_{i,j}[n]$ looks like in various examples.

\begin{figure}[h]
\centering
\includegraphics[width=6in]{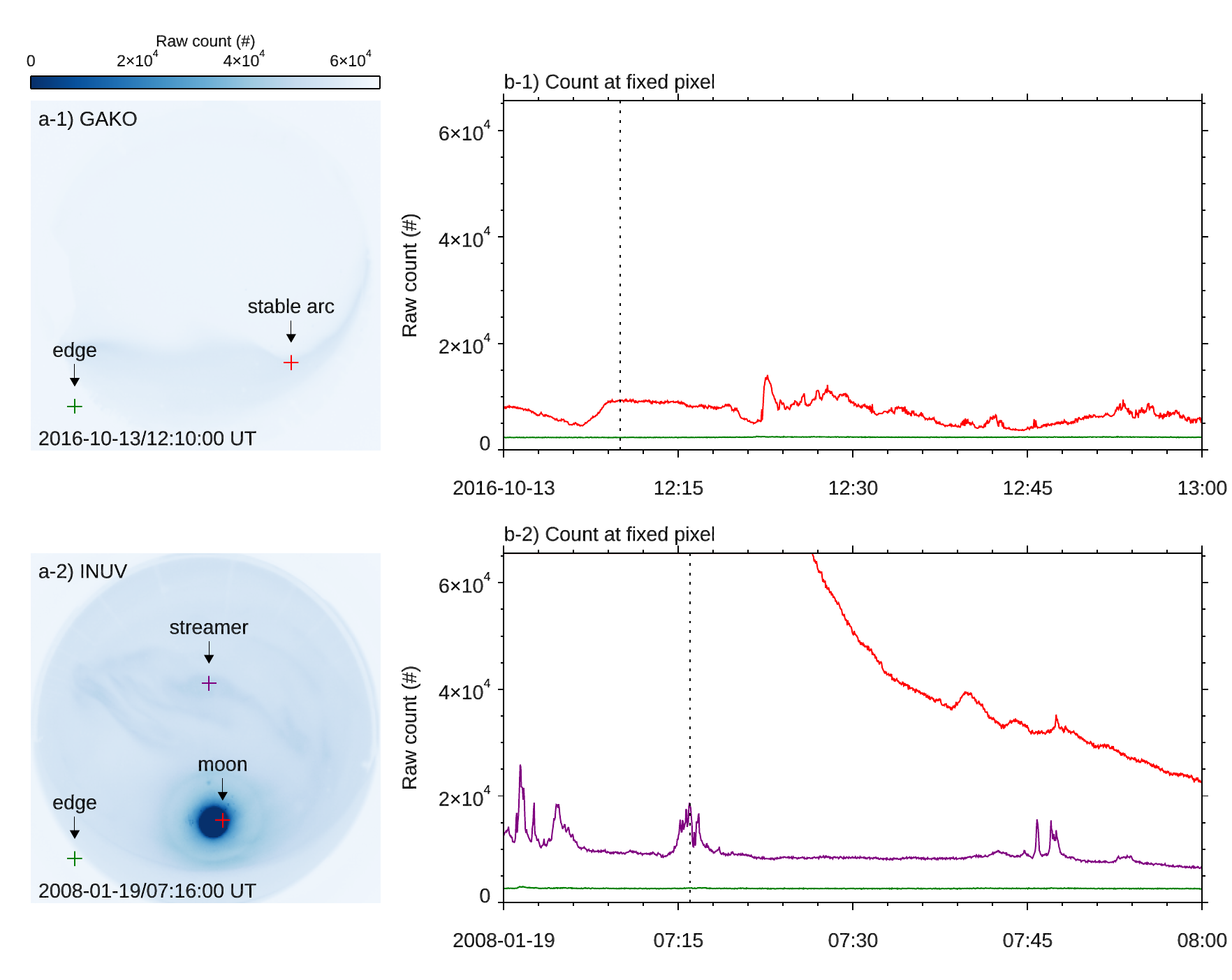}
\caption{Examples of the ASI images (Panels a-1 and a-2) at selected times and the time series of the count at selected pixels (Panels b-1 and b-2). The $+$ symbols in Panels a-1 and a-2 mark the position of the selected pixels. The dashed lines in Panels b-1 and b-2 mark the selected times. }
\label{fig:raw_count}
\end{figure}

In the first example, the snapshot (Panel a-1) of GAKO at 2015-10-13/12:10 UT shows that a stable arc was passing the pixel at the red symbol. The count at the red pixel is shown in Panel b-1. The plateau from 12:06 to 12:21 UT is due to the stable arc. As a comparison, a green symbol marks the edge of the field of view in Panel a-1. The count at the green pixel was stable during the 1 hour of data. In the second example, the snapshot (Panel a-2) of INUV at 2006-01-19/07:16 UT shows that the camera captured several streamers and the moon. Here, we selected a sample pixel (red) for the moon and another sample pixel (purple) for one of the streamers. As shown in Panel b-2, the moon saturated the count of the red pixel ($=65535$) before 07:26 UT. After this time, the moon gradually moved away from the red pixel, causing the count to gradually decrease. In contrast, the count of the purple pixel contains several spikes, due to the passage of streamers.

From these examples, we can see that $r_{i,j}[n]$ contains the information on where it is:
\begin{itemize}
\item For a pixel on the edge, $r_{i,j}[n]$ is almost a constant and the raw count is very low
\item For a pixel passed by auroral forms (stable arc, streamer, etc), $r_{i,j}[n]$ remains around a stable baseline level but can increases on the order of $1\times 10^4$. These increases can be short (3-5 min for streamers) or long (10-20 min for stable arcs). Note that clouds can also cause similar count increases as the aforementioned auroral forms. Thus clouds cannot be treated by the algorithm.
\item For a pixel passed by the moon, $r_{i,j}[n]$ is dominated by a significant increase (or decrease) of the baseline level. It can reach the saturation count (65,535) and can contain the normal count increase on the order of $1\times 10^4$ due to auroral forms.
\end{itemize}

Based on these observations, we can see that the key is to properly calculate the baseline level for $c_{i,j}[n]$, i.e., the "background" count $b_{i,j}[n]$. In the later sections, we will explain the key technical details and conclude with the recipe and IDL program of the algorithm.

\begin{figure}[h]
\centering
\includegraphics[width=5.5in]{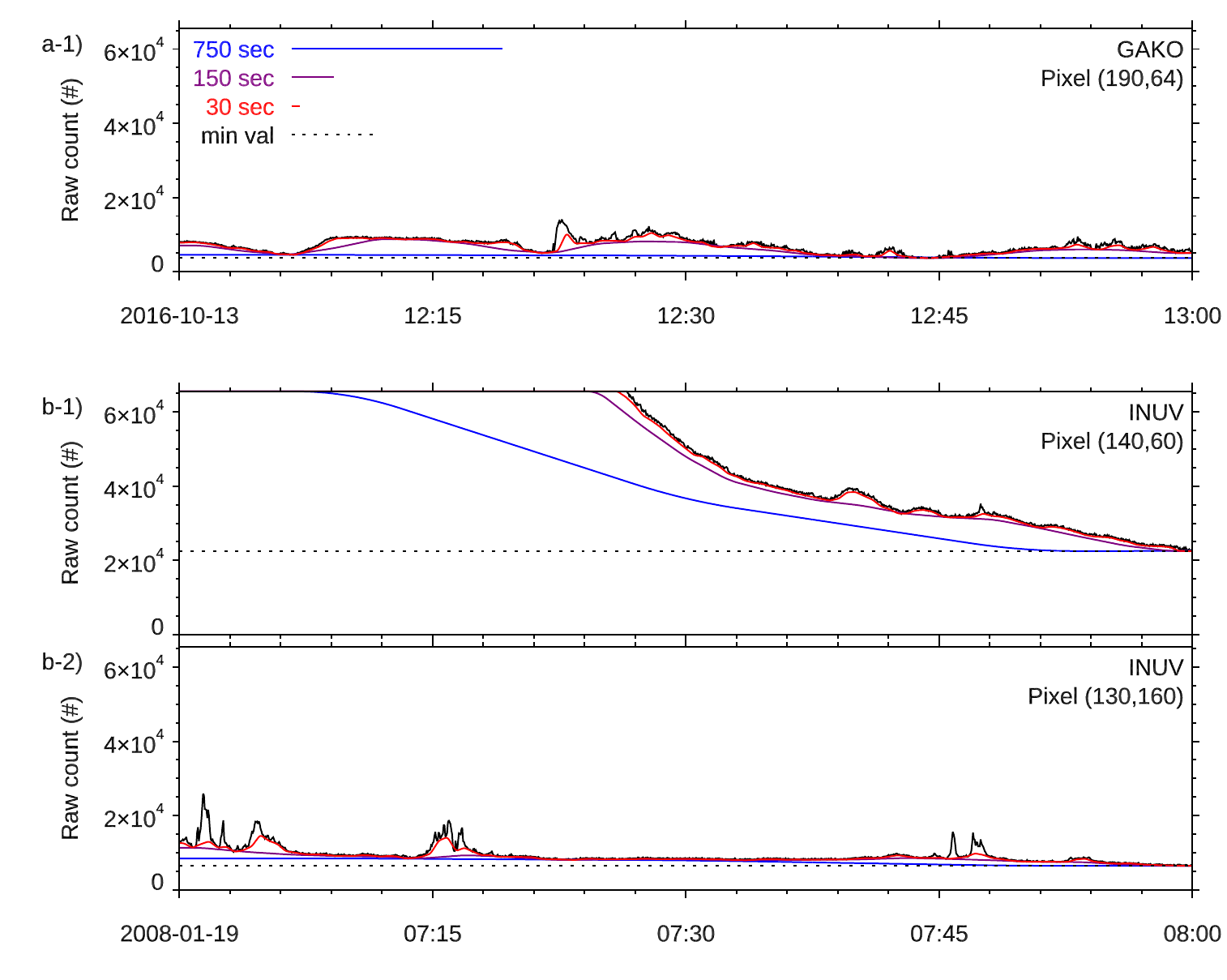}
\caption{The count of selected pixels and the baselines calculated from different window sizes. Panel a-1 shows the example at GAKO and Panels b-1 and b-2 show the example at INUV. All these examples have been shown in Fig \ref{fig:raw_count}. In each panel, baselines for different window sizes are calculated as explained in Section \ref{sec:low_env}. The dashed line marks the minimum count. }
\label{fig:low_env}
\end{figure}

\section{The Baseline of Count Variation per Pixel}
\label{sec:low_env}

Fig \ref{fig:low_env} shows the time series of the count at selected pixels shown in Fig \ref{fig:raw_count}. We can see that for the examples in Panels a-1 and b-2, the minimum value (dashed line) is already a good estimate of the baseline of $r_{i,j}[n]$ (black). The blue, purple, and red lines are 3 estimates of the baselines at progressively smaller window size. The calculation of the baseline is listed in Appendix \ref{sec:code}. The essential idea is to truncate the time series into several sectors at the specified window size. We then get the minimum value within each sector and use these minimum values to form the baseline. Therefore, decreasing the window size causes the baseline to move closer to $r_{i,j}[n]$. In the extreme case, if the window size is 3 sec, the baseline is exactly $r_{i,j}[n]$, of if the window size is $n\times 3$ sec, then the baseline is the minimum value. Here we define the window $w$ in sec. $w$ equals to the window size divided by 3 sec. We use $\beta^w_{i,j}[n]$ to refer to the baseline for $r_{i,j}[n]$ for a given window $w$.

From the panels in Fig \ref{fig:low_env}, we can see that for the case of moon (Panel b-1), we need a small window to follow the count decreasing caused by the moon. The baselines at large width (e.g. $w=750$ sec) cannot capture the count change due to the moon. However, for the cases of auroral forms (Panels a-1 and b-2), we need a large window. If the width is too small ($w=30$ sec), the baseline will capture the change due to auroral forms. This causes them to be removed because we essentially want $r_{i,j}[n]-\beta_{i,j}^w[n]$ to get the calibrated ASI data.

\begin{figure}[h]
    \centering
    \includegraphics[width=4.5in]{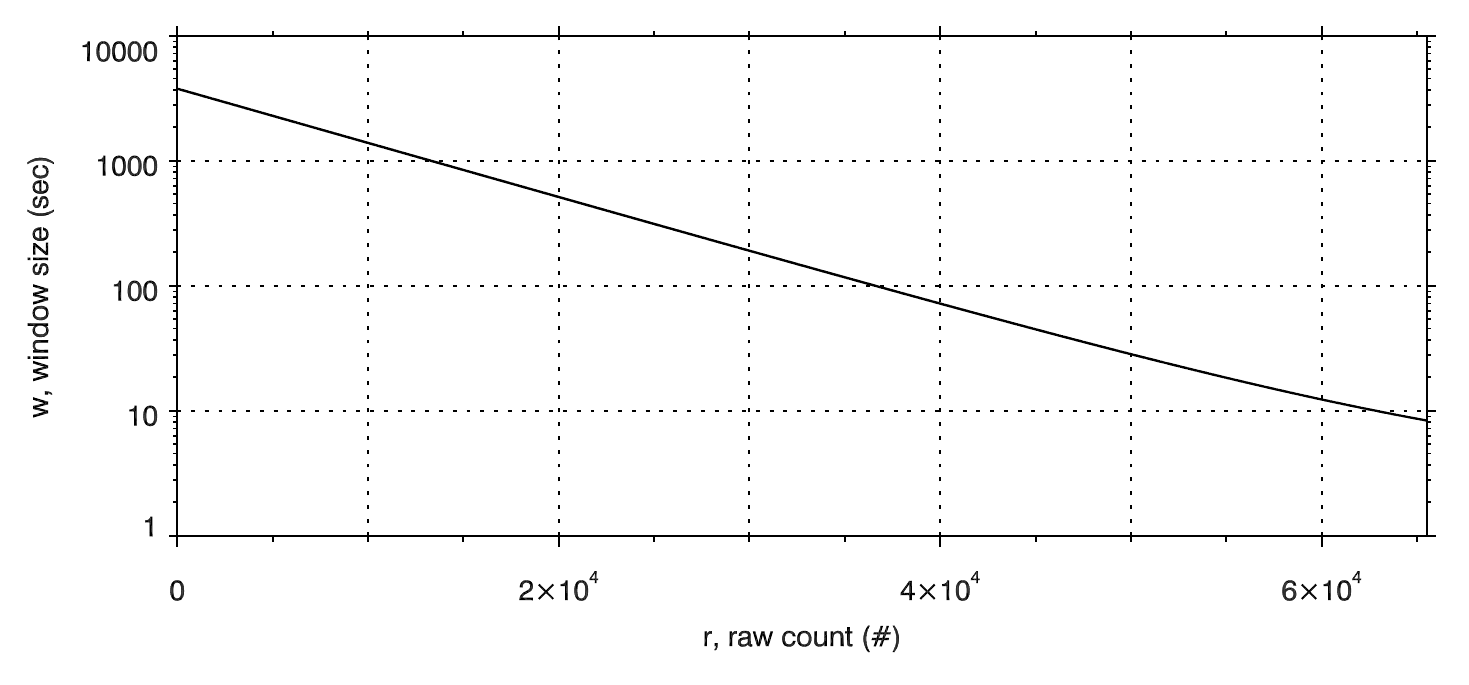}
    \caption{The mapping from count to width, as defined by Equation (\ref{eq:map}).}
    \label{fig:map}
\end{figure}

\subsection{Moon weight}
Because the moon's surface and its glow correspond to high raw count. We weight the raw count $r_{i,j}[n]$ with the distance between the moon and each pixel. In the algorithm, we first calculate $\theta_{i,j}[n]$, which is the angle between each pixel and the center of the moon, and then calculate the moon weight $m_{i,j}[n]$ from Equation (\ref{eq:moon}).
\begin{equation}
    m_{i,j}[n] = 1+2\exp^{1-\theta_{i,j}[n]/2.5}.
    \label{eq:moon}
\end{equation}
From the equation, we can see that the moon weight is 2-3 when close to the moon ($<2.5$ deg) decays to 1 when far away from the moon. We weight the raw count by the moon weight and define the weighted count $r'_{i,j}[n] = r_{i,j}[n]\times m_{i,j}[n]$. The weighted count $r'$ will be used to calculate the adaptive width. Note that we do not use the moon's position to remove the moon directly because there is also the moon's glow, whose size varies as weather changes. The moon weight system is adopted to treat both the moon's surface and glow.

\subsection{Adaptive window size}
As shown in Fig \ref{fig:low_env}, we need a mixture of the baselines at different windows, where smaller window size is needed for higher raw count. For example, we need a small window for the moon and a large window for auroral arcs. Based on tests on the ASI images, we adopt the following empirical formula for mapping counts to window:
\begin{equation}
w = 3+2\exp^{5.5655-r'\times10^{-4}},
\label{eq:map}
\end{equation}
where $r'$ is the weighted count. Equation (\ref{eq:map}) is visualized in Fig \ref{fig:map}. The window is $>1200$ sec (20 min) for $r'<10^4$, which is essentially a constant for such counts, to preserve the count changes due to auroral structures. The window is much smaller ($<100$ sec) for large counts ($r'>4\times 10^4$), to capture the count changes due to the moon and its glow.

\section{Calculate the Background}
\label{sec:calc_bg}

Ideally, for a given time series $r_{i,j}[n]$, we calculate its baseline at an adaptive width $w_[i,j][n]$ which is determined from $r_{i,j}[n]$ using Equation (\ref{eq:map}). However, this is too computationally expensive because it loops through all iterators $i,j,k$. In practice, our algorithm does not directly calculate the adaptive baseline $\beta_{i,j}[n]$. Instead, we estimate it by interpolating among the baseline calculated at several fixed sampling widths. For the time series $r_{i,j}[n]$, this estimation is the "background" we will use and is referred to as $b_{i,j}[n]$. Looping through all pixels results in the overall background $b_{m,m}[n]$. Its difference from the raw images gives us the calibrated images $c_{m,m}[n] = r_{m,m}[n]-b_{m,m}[n]$.

As shown in Fig \ref{fig:low_env}, the baseline approaches $r_{i,j}[n]$ as the window size is decreased. In our algorithm, we choose the following sampling windows at $[1800,180,3]$ sec. They are referred to as $w_s$, where $s=0,1,2$. For a given $r_{i,j}[n]$, we actually only calculate $\beta^{w_1}_{i,j}[n]$ for the width $w_1 =180$ sec. For $w_0=1800$ sec, we set $\beta_{i,j}^{w_0}[n] \equiv \min({r}_{i,j}[n])$ (a constant) and for $w_2 = 3$ sec, we set $\beta_{i,j}^{w_2} = r_{i,j}[n]$ (the time series itself). With these baselines at the fixed sampling widths, for a given pixel $r_{i,j}[k]$, we have an array $\beta_{i,j}^{w_s}[k]$ depending on the iterator $s$. We also have its adaptive width $w_{i,j}[k]$ calculated from Equation (\ref{eq:map}). Therefore, we can estimate the adaptive baseline as 
\begin{equation}
b_{i,j}[k] = interpol(\beta_{i,j}^{w_s}[k], w_s, w_{i,j}[k]).
\label{eq:bg}
\end{equation}
Looping though all pixels in $i$, $j$, and $k$ results in the background $b_{m,m}[n]$.

\begin{figure}[h!]
\centering
\includegraphics[width=5in]{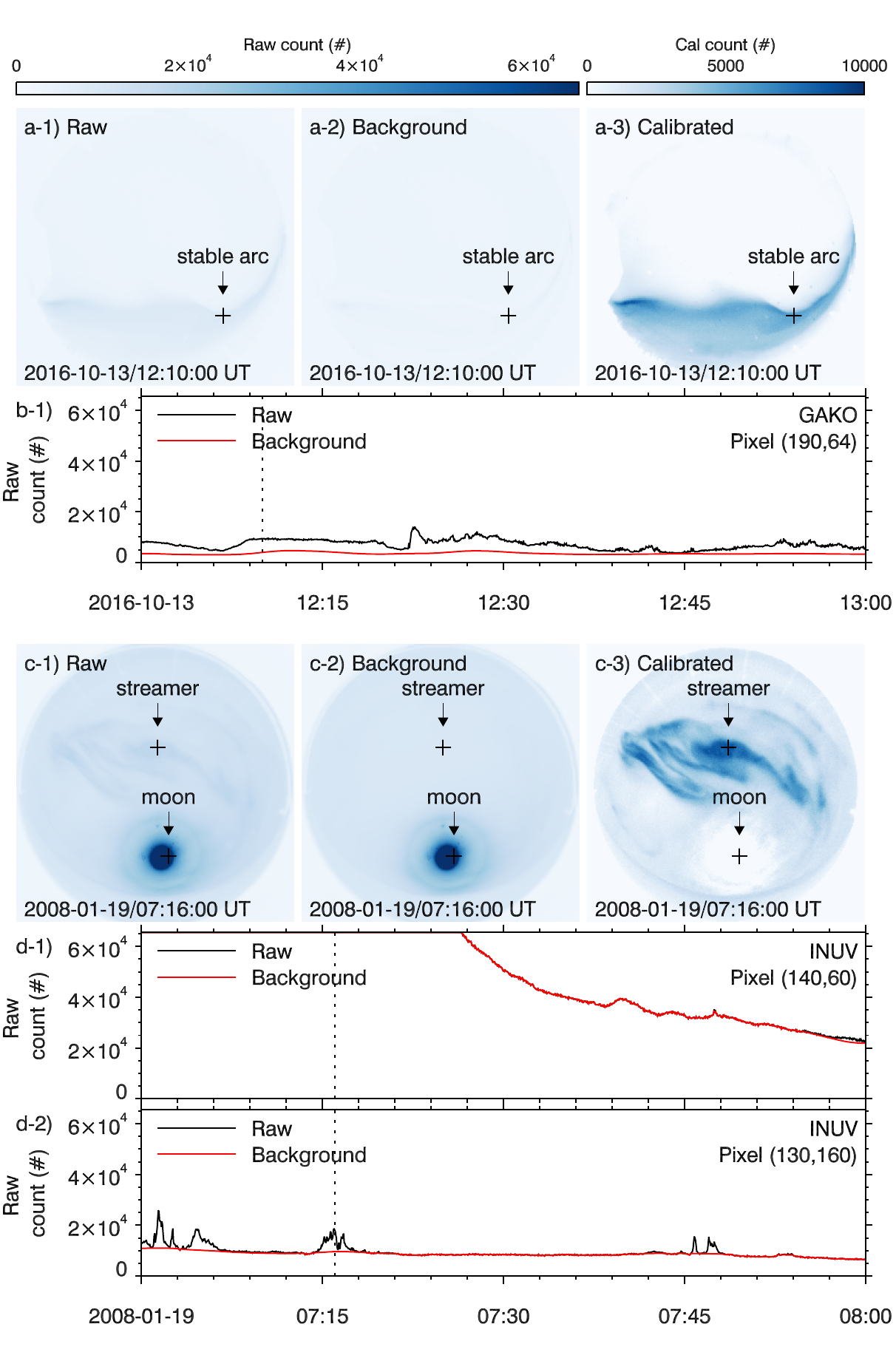}
\caption{Examples showing the raw, background, and calibrated images and the time series of the count at selected pixels.}
\label{fig:old_vs_new}
\end{figure}

Fig \ref{fig:old_vs_new} shows the raw, background, and calibrated images for the same examples in Fig \ref{fig:raw_count}. In the first example, Panel a-1 shows the raw image $r_{m,m}[k]$ at frame $k$. Panel a-2 shows the background $b_{m,m}[k]$ and Panel a-3 shows the calibrated image $c_{m,m}[k]$. Panel b-1 shows the time series $r_{i,j}[n]$ and the background $b_{i,j}[n]$ at the selected pixel. We can see that the stable arc is preserved in Panel a-3. The background in Panel b-1 is close to a flat background as desired.

In the second example shown in Panels c and d. The calibrated image in Panel c-3 preserves the streamer and the moon and its glow are removed. In Panel d-1, the background is very close to the decreasing trend caused by the moon. In Panel d-2, the background does not follow the peaks related to the streamer. Both are as desired.

Note that among the defined baselines, $\beta_{i,j}^{w_1}[n]$ may be inaccurate for the times on edges, i.e., those data within the first and last $w_1=180$ s windows. For example, in Panel d-1, within the last 3 min, the background deviates slightly more from the raw count than previous times. This can be fixed by padding the data by 180 s at both the start and end. Here, this is deliberately skipped to demonstrate the deviation. Such padding is highly recommended under essentially all circumstances.

\section{Conclusions and discussion}
\label{sec:conclusion}

We have discussed a new algorithm to remove the moon from the raw ASI images. The idea is to analyze the change of counts at each pixel and remove a baseline background in an adaptive manner. We have tested the algorithm extensively on the THEMIS ASI data. Based on these tests, the algorithm can effectively remove the moon and recover auroral structures within the moon's glow.

One known limitation of the algorithm is that clouds are not treated. In addition, we note that the performance of the algorithm is primarily determined by how the moon weight, adaptive window, and background are defined. We believe that the current definitions for these parameters (Equations (\ref{eq:moon}), (\ref{eq:map}), and (\ref{eq:bg})) results in satisfactory results in general for the tested THEIS ASI data. However, further optimization on these definitions is very likely to further improve the performance of the algorithm. Here, the main purpose of this document is to explain the essence of the algorithm, i.e., why we introduce these parameters instead of how they are exactly defined. For general ASI data, our algorithm should work as well, but with modified definitions to accommodate the dynamic range of the target ASI data.

\newpage

\newpage
\appendix
\section{Programs}
\label{sec:code}
This section lists the IDL programs for implementing the algorithm. These programs are accessible at Github at \url{https://github.com/tsssss/slib/tree/master/sread/themis/utils}. We may update the programs in the future. Thus please check the Github URL for the latest version.
\subsection{Calculate baseline}
\begin{lstlisting}[language=idl,numbers=left,basicstyle=\ttfamily\scriptsize]
function calc_baseline, asi_raw, window, times

    ; Truncate data into sectors of the wanted window
    nframe = n_elements(asi_raw)
    sec_times = make_bins(minmax(times), window, inner=1)
    time_step = 3
    sec_pos = (sec_times-times[0])/time_step
    nsec = n_elements(sec_pos)-1
    frames = dindgen(nframe)

    ; Get the min value within each sector.
    xxs = fltarr(nsec)
    yys = fltarr(nsec)
    for kk=0,nsec-1 do begin
        yys[kk] = min(asi_raw[sec_pos[kk]:sec_pos[kk+1]-1], index)
        ;xxs[kk] = sec_pos[kk]+index  ; This causes weird result.
        xxs[kk] = (sec_pos[kk]+sec_pos[kk+1])*0.5
    endfor
;    ; Add sample points at the beginning and end of the raw data.
    txs = frames[sec_pos[0]:sec_pos[1]]
    tys = asi_raw[sec_pos[0]:sec_pos[1]]
    res = linfit(txs,tys)
    ty = (yys[0]-(xxs[0]-0)*res[1])>min(tys)
    xxs = [0,xxs]
    yys = [ty,yys]
    
    txs = frames[sec_pos[-2]:sec_pos[-1]]
    tys = asi_raw[sec_pos[-2]:sec_pos[-1]]
    res = linfit(txs,tys)
    ty = (yys[-1]+(nframe-1-xxs[-1])*res[1])>min(tys)

    ; Smooth after interpolation to make the background continuous.
    time_bg = smooth(interpol(yys,xxs,frames), window*0.5, edge_mirror=1)
    return, time_bg

end
\end{lstlisting}

\subsection{Calibrate ASI images}
\begin{lstlisting}[language=idl,numbers=left,basicstyle=\ttfamily\scriptsize]
pro themis_asi_cal_brightness, asf_var, newname=newname

    if n_elements(newname) eq 0 then newname = asf_var+'_norm'
    get_data, asf_var, times, imgs_raw, limits=lim
    image_size = double(size(reform(imgs_raw[0,*,*]), dimensions=1))
    nframe = n_elements(times)
    time_step = 3d

;---Calculate the background of edge pixels.    
    deg = constant('deg')
    rad = constant('rad')
    pixel_elevs = lim.pixel_elev*rad
    pixel_azims = lim.pixel_azim*rad

    edge_indices = where(finite(pixel_elevs,nan=1) or pixel_elevs le 0, nedge_index, $
        complement=center_indices, ncomplement=ncenter_index)
    center_indices_2d = fltarr(ncenter_index,2)
    center_indices_2d[*,0] = center_indices mod image_size[0]
    center_indices_2d[*,1] = (center_indices-center_indices_2d[*,0])/image_size[0]
    

;---A minimum background per pixel.
    imgs_bg = fltarr([nframe])
    tmp = reform(imgs_raw,[nframe,product(image_size)])
    for ii=0,nframe-1 do begin
        imgs_bg[ii] = min(tmp[ii,center_indices])
    endfor
    width_slow = 60
    imgs_bg = smooth(imgs_bg,width_slow, nan=1, edge_mirror=1)
    imgs_bg0 = fltarr([nframe,image_size])
    for ii=0,nframe-1 do imgs_bg0[ii,*,*] = imgs_bg[ii]

    
;---Remove fast varying signals.
    window_slow = width_slow*time_step
    imgs_slow = imgs_raw
    for index_id=0,ncenter_index-1 do begin
        ii = center_indices_2d[index_id,0]
        jj = center_indices_2d[index_id,1]
        imgs_slow[*,ii,jj] = calc_baseline(imgs_slow[*,ii,jj], window_slow, times)
    end
    
    
;---Prepare adpative window.
    sample_windows = [1,width_slow,600]*3
    nsample_window = n_elements(sample_windows)
    
    
;---Calculate the moon's elevation and azimuth.
    site_glon = lim.asc_glon
    site_glat = lim.asc_glat
    moon_elevs = moon_elev(times, site_glon, site_glat, $
        azimuth=moon_azims)
    moon_xpos = cos(moon_elevs)*cos(moon_azims)
    moon_ypos = cos(moon_elevs)*sin(moon_azims)
    moon_zpos = sin(moon_elevs)


;---Calculate the angle between each pixel and the moon.
    moon_angles = fltarr([nframe,product(image_size)])
    foreach center_index, center_indices do begin
        pixel_elev = pixel_elevs[center_index]
        pixel_azim = pixel_azims[center_index]
        pixel_xpos = cos(pixel_elev)*cos(pixel_azim)
        pixel_ypos = cos(pixel_elev)*sin(pixel_azim)
        pixel_zpos = sin(pixel_elev)
        moon_angles[*,center_index] = acos(moon_xpos*pixel_xpos+moon_ypos*pixel_ypos+moon_zpos*pixel_zpos)
    endforeach
    moon_angles = reform(moon_angles,[nframe,image_size])*deg
    moon_weight = exp((2.5-moon_angles)/2.5)*2+1 ; decays to 1.

    
;---Calulate the background of imgs_slow.
    max_count = 65535
    norm_count = round(imgs_slow*moon_weight)<max_count
    norm_windows = exp((max_count-norm_count+1e4)/1e4)*2+3
    
    index = where(norm_windows ge sample_windows[1], count, complement=index2, ncomplement=count2)
    imgs_bg = fltarr([nframe,image_size])
    if count ne 0 then begin
        weights = (norm_windows[index]-sample_windows[1])/(sample_windows[2]-sample_windows[1])
        imgs_bg[index] = weights*imgs_bg0[index]+(1-weights)*imgs_slow[index]
    endif
    if count2 ne 0 then begin
        weights = (norm_windows[index2]-sample_windows[0])/(sample_windows[1]-sample_windows[0])
        imgs_bg[index2] = weights*imgs_slow[index2]+(1-weights)*imgs_raw[index2]
    endif
    
    imgs_cal = imgs_raw-imgs_bg
    imgs_cal = reform(imgs_cal,[nframe,product(image_size)])
    bg0 = fltarr(nframe)
    for kk=0,nframe-1 do begin
        index = where(moon_angles[kk,*,*] ge 40, count)
        tmp = imgs_cal[kk,index]
        tmp = tmp[sort(tmp)]
        bg0[kk] = tmp[0.1*count]
    endfor
    bg0 = smooth(bg0,width_slow,nan=1,edge_mirror=1)
    imgs_cal = reform(imgs_cal,[nframe,image_size])
    for kk=0,nframe-1 do begin
        imgs_cal[kk,*,*] -= bg0[kk]
    endfor
    imgs_cal >= 0
    
    store_data, newname, times, imgs_cal, limits=lim
end
\end{lstlisting}


\end{document}